# Abrupt efficiency collapse in real-world complex weighted networks: robustness decrease with link weights heterogeneity

Bellingeri M.[1*], Bevacqua D.[2], Scotognella F.[3,4], Cassi D.[1]

[1]Dipartimento di Fisica, Università di Parma, via G.P. Usberti, 7/a, 43124 Parma, Italy

* Corresponding author: michele.bellingeri@unipr.it

[2]PSH, UR 1115, INRA, 84000, Avignon, France

[3]Dipartimento di Fisica, Istituto di Fotonica e Nanotecnologie CNR, Politecnico di Milano, Piazza Leonardo da Vinci 32, 20133 Milano, Italy

[4]Center for Nano Science and Technology@PoliMi, Istituto Italiano di Tecnologia, Via Giovanni Pascoli, 70/3, 20133, Milan, Italy

**Here we report a comprehensive analysis of the robustness of five high-quality real-world complex weighted networks to errors and attacks of nodes and links. We analyze *C. Elegans*, Cargo ship, *E. Coli*, Us Airports and Human brain real-world complex weighted networks. We use measures of the network damage conceived for a binary (e.g. largest connected cluster *LCC*, and binary efficiency $Eff_{bin}$) or a weighted network structure (e.g. the efficiency *Eff*, and the total energy-information *En*). We find that removing a very small fraction of nodes and links with respectively higher strength and weight triggers an abrupt collapse of the weighted functioning measures while measures that evaluate the binary-topological connectedness are almost unaffected. These findings unveil a problematic response-state of the real-world complex networks where the attack of a small fraction of nodes-links returns these systems in a *connected but highly inefficient state*. Our findings unveil how the robustness may be heavily overestimated focusing on the connectedness of the components only. Last, to understand how the networks robustness is generally affected by link weights heterogeneity, we randomly**

**assign link weights over the topological structure of the real-world networks. We find that highly heterogeneous networks experienced a faster efficiency decrease under nodes-links removal: i.e. the robustness of the real-world complex networks against both random than attack is negatively correlated with links weight heterogeneity.**

The robustness of a network is its ability to maintain the system functioning in case of failures of nodes or links. Networks robustness is extremely important and has been widely investigated in last years in different fields of science[1-11]. A comprehensive analysis of network robustness considers the failure of both nodes (e.g. Iyer et al.[8]) and links (e.g. Pajevic and Plenz[12]). Initially, large attention has been dedicated to binary-topological analyses. Yet, recent studies evidenced that the robustness of complex networks can be comprehensively understood only when considering the strength (weight) of the relationship (links) among nodes[12-22].

The analysis of the robustness of complex weighted networks provided fundamental outcomes. Past studies demonstrated that when network connectivity is measured using the largest connected cluster (*LCC*), it is highly vulnerable to the removal of links with lower weight (weak links) but robust to deletion of links of higher weight (strong links)[17-20]. The final outcome was that 'weak links are the universal key for complex networks stability'[20]. On the other hand, Pajevic and Plenz[12] outline how the average clustering of nodes (that can be viewed as a measure of the local efficiency of the system) is robust to the removal of weak links but rapidly destroyed when removing links with higher strength. Dall'Asta et al.[21] showed that introducing the weight of links in the US airports network would decrease its robustness with respect to classic topological frameworks. Further, Bellingeri and Cassi[22] outlined how the network robustness response to node attacks changes according to the considered measures of the system functioning, i.e. weighted or binary.

In the present work, we investigate the role of the weighted structure of complex networks in shaping their robustness against both nodes and links failure. We analyze a high quality set of real-world weighted complex networks from different fields of science (Table S1). The considered network present different number of nodes, links, and a sound interpretation of the nature of link weights, e.g. in the US airports network, the weight identifies the passengers flowing from two airports; in the brain network of the nematode *C. Elegans*, it identifies the number of connections joining neurons. We randomly removed nodes or links to simulate an error in the system, and we eliminated nodes with higher number of links and with higher strength (e.g. higher sum of link weights) and links with higher weight (Supplementary materials S2) to simulate an attack. This is the so-called *attack strategy* with nodes or links removed according to some structural properties of the network [1-9]. We evaluated the robustness of the complex networks to nodes or links in terms of the decrease of network functioning measures reflecting both the binary-topological and the complex weighted structure of the system[3,7,8,13]. The *LCC*[1-3] is the largest number of nodes connected by at least one path in the network and can be viewed as a binary (unweighted) measure of the network functioning. The efficiency[13-14,22] (*Eff*) is a measure of the global complex network capacity to deliver information among system constituents and allows a precise quantitative evaluation of unweighted ($Eff_{bin}$) and weighted networks functioning (*Eff*). A decrease in the efficiency means a reduction in the energy-information pace exchanging over the network. The total energy-information flowing in the system[21] (*En*) is the sum of the links weights; it represents the simplest weighted measure evaluating the actual flows in the networks (Supplementary Materials S3). When links or nodes are removed from the network we can assess the decrease of the system functioning according to different measures as showed in Fig. 1a. The more important are the components removed from the network, the steeper is the decrease in the network functioning measure. For example, in Figure 1a the red removal strategy identifies more important components in the network, since a given fraction *q* of nodes-links removal, triggers a steeper decrease in the network functioning efficiency (*Eff*, normalized on the initial maximal value) with respect to the black strategy. To compare the response among networks and measures, we resume the removal outcomes in a single value defined as the network robustness (*R*), reported in Figure 1b. The robustness *R* corresponds to the area below the curve of the system functioning against the fraction of nodes-links removed and ranging between two theoretical extremes, $R \cong 0$ and $R=1$.

In Fig. 1 we show the robustness (*R*) outcomes under different types of nodes-links removals. When quantifying the system functioning with *Eff* we find the real-world complex networks to be highly vulnerable to the removal of links with higher weights (Figure 1c); i.e. *Strong* strategy produces the fastest decrease of the system efficiency functioning (*Eff*). Further, we found real-world complex networks *Eff* to be very robust to the deletion of weaker links and *Weak* links attack strategy is highly ineffective even causing an efficiency decrease lower than the random removal of links (*Rand*). At the opposite, when measuring the network functioning with *LCC*, we find *Strong* and *Weak* links attack strategies the less and the most effective to reduce the *LCC* (Fig. 1c). This confirm recent analyses showing that the deletion of strong links preserves the *LCC* until the weakest links are removed [17-20].

In real-world networks, link weights are coupled to the binary topology in a non-trivial way[17-20-21,23], for instance with nodes strength-degree correlation meaning that links with higher weight are more probable joining high degree nodes[14,15]. For this reason it is important to understand why strong links are important in supporting *Eff* because to their large weight or for their specific embedding among more connected nodes (hubs) of the network. We compute the robustness of the real-world networks after weights randomization over the topological structure, with link weights independent of any topological features and acting as a control outcome. *Strong* strategy results to be the most harmful to decrease *Eff* even when randomizing the real weights over the binary-topological structure, i.e. the importance of strong links to support *Eff* is maintained also when the real weight-topology coupling is no correlated (Fig. 1a, see *Eff$_{ran}$*). We perform an *experimentum crucis* removing links according to the real link weights, then measuring *Eff$_{bin}$* considering links like binary (all weights equals to 1). In this way we nullify the influence of the weight to shape the efficiency (*Eff*) maintaining only the binary-topological role of the strong links. We discover an *efficiency reversal pattern* for all the real-world complex networks and now weak links removal readily decrease the efficiency functioning whereas *Strong* become ineffective (Figure 2a, *Eff-Eff$_{bin}$* column). In other words, the importance of strong links to support the information delivery efficiency is mainly due to their larger intensity, with a secondary role of their topological embedding. These results bring important evidences inside the long standing debate about the importance of weak and strong links[17-21,24] showing that links carrying larger weight would be fundamental to support the efficiency of the system hence not being responsible of the topological connectedness of the network. We also revise the importance of weak links in support network robustness confirming their function in maintaining the topological connectedness of the network[17-20] but also their small relevance for information delivery efficiency (Figs 1c, 2a). Very important, we outline that removing a small fraction of strong links can readily collapse the real-world network information delivery efficiency(*Eff*) despite the size of the largest connected cluster (*LCC*) is still preserved. We illustrate this finding in Fig. 3 a,b. For example, removing a small fraction of busy shipping routes (10% strong links) from Cargo-ship network produced a quick collapse in the system efficiency (50% *Eff*) isolating only 2% of ports-nodes (2% *LCC*) (Fig. S7).

We find that real-world networks are robust to the random removal but vulnerable to the deletion of higher connected and higher strength nodes for all the functioning measures, i.e. real-world weighted networks would be "*random resistance*" and "*attack prone*" (Fig. 1d) confirming classical binary-topological outcomes[1-9]. Very interesting, removing a handful of nodes abruptly collapse the functioning efficiency (*Eff*) of the networks. Removing only 3 highly strength nodes-neurons in the *C. Elegans* network produces the sharp collapse to 50% of the initial value of the *Eff* whereas the *LCC* is roughly totally preserved (Fig. 3d-f); removing 4 nodes-metabolites among the 1100 total nodes-metabolites in *E. Coli* network decrease the 30% of the efficiency and only the 2% the *LCC*; in the US Airports network, 5 nodes-airports removals over the N=500 airports sharply decrease the total energy-information (*En*) to the 60% of the initial value with 5% *LCC* decrease (Fig. S8); 5 nodes-ports attack reduces to the 70% the efficiency in the Cargo-ship network. In all the networks the binary efficiency (*Eff$_{bin}$*) follows the *LCC* with a very slow decrease. Only for the Human brain network we find a small difference in the measurements with *Eff* closed to the *LCC* (Fig. S8). These outcomes can be resumed in a novel and problematic response pattern of the real-world complex networks, where the removal of nodes-links components playing a major role in the energy-information delivery may leave the system in a *connected but highly inefficient state* (Fig. 3).

Dall'Asta et al.[21] showed that when removing highly connected nodes the total energy-information (that the authors called 'outreach') of the US Airports network decreased much rapidly than its *LCC* measure. Our findings wide Dall'Asta et al.[21] outcomes for different kinds of network and measurements, unveiling that even removing handful nodes can leave real-world complex weighted networks in a *connected* (the *LCC* and *Eff$_{bin}$* are preserved) *but highly inefficient state* (*Eff* and *En* quickly collapse). This evidence outlines how using binary measurements may heavily overestimate the robustness of real-world networks.

Very interesting, we find that real-world networks exhibit higher efficiency (*Eff*) robustness to the removal of nodes after the link weight randomization (except the Human brain), both random than attack (Fig. 2b, *Eff-Eff$_{ran}$*, and Fig. S6, S10). In networks with strength-degree correlation, the removal of higher connected nodes (the so-called hubs) will delete strongest links with higher energy-information loss; differently, in the control network, the weights randomization eliminates the correlation and the removal of hubs would intercept less energy-information. This finding indicates that some level of nodes degree-strength coupling discovered in real-world complex networks[15-16,21,24] would make these systems even more vulnerable to nodes removal.

To examine in depth how link weights pattern influences the robustness of real-world networks we removed nodes-links with artificially increasing the link weights heterogeneity, by assigning links weight sorted from rectangular and from a 2 values distribution (Supplementary materials S4). The weights heterogeneity is enhanced by progressively increasing the maximum weight ($Wmax$). Surprisingly, increasing $Wmax$ we assist to a decrease in robustness ($Eff$) for all the nodes removal strategies (Fig. 4a) whereas the proportion of energy-information ($En$) subtracted to the networks remains roughly constant (Fig. 4a). For random links removal we find the same pattern (Fig. 4b). All these trends are more pronounced for the 2 values distribution of link weights (Fig. S11). This would indicate that the decrease in robustness efficiency is not due to a major amount of 'weight' intercepted in the networks with increasing $Wmax$, but it would be an effect of the larger link weights heterogeneity. Differently the energy-information subtracted in the network by *Strong* and *Weak* links removal is clearly related to the weights heterogeneity, and we observe a robustness decrease for strongest (and an increase for weakest) links removal for both $En$ and $Eff$ measures (Fig. 4b).

The last finding showed that the robustness decrease is not only related to the transition from binary ($LCC$) to weighted measurement ($En$) as suggested from Dall'Asta et al.[21], but is within a more general mechanism by which enhancing links weight heterogeneity negatively affects the robustness ($Eff$) in real-world complex networks (Fig. 4, Fig. S11). The heterogeneity in link weights is interpreted as a feature able to stabilize different real-world networks[20] with these systems self-organizing toward large heterogeneity in links weight (with many weak links). Our discoveries would indicate that if the complex networks self-organizing with large weight heterogeneity, they pay the price in terms of robustness, with potential higher vulnerability to nodes-links failure.

**Acknowledgement**

Many thanks to E. Agliari for a seminal suggestion on the main idea of this paper. We thank S. Vincenzi for constructive comments during this work. We thank A. Allard and M. Boguna for sharing their real-world networks dataset.

**FIGURES**

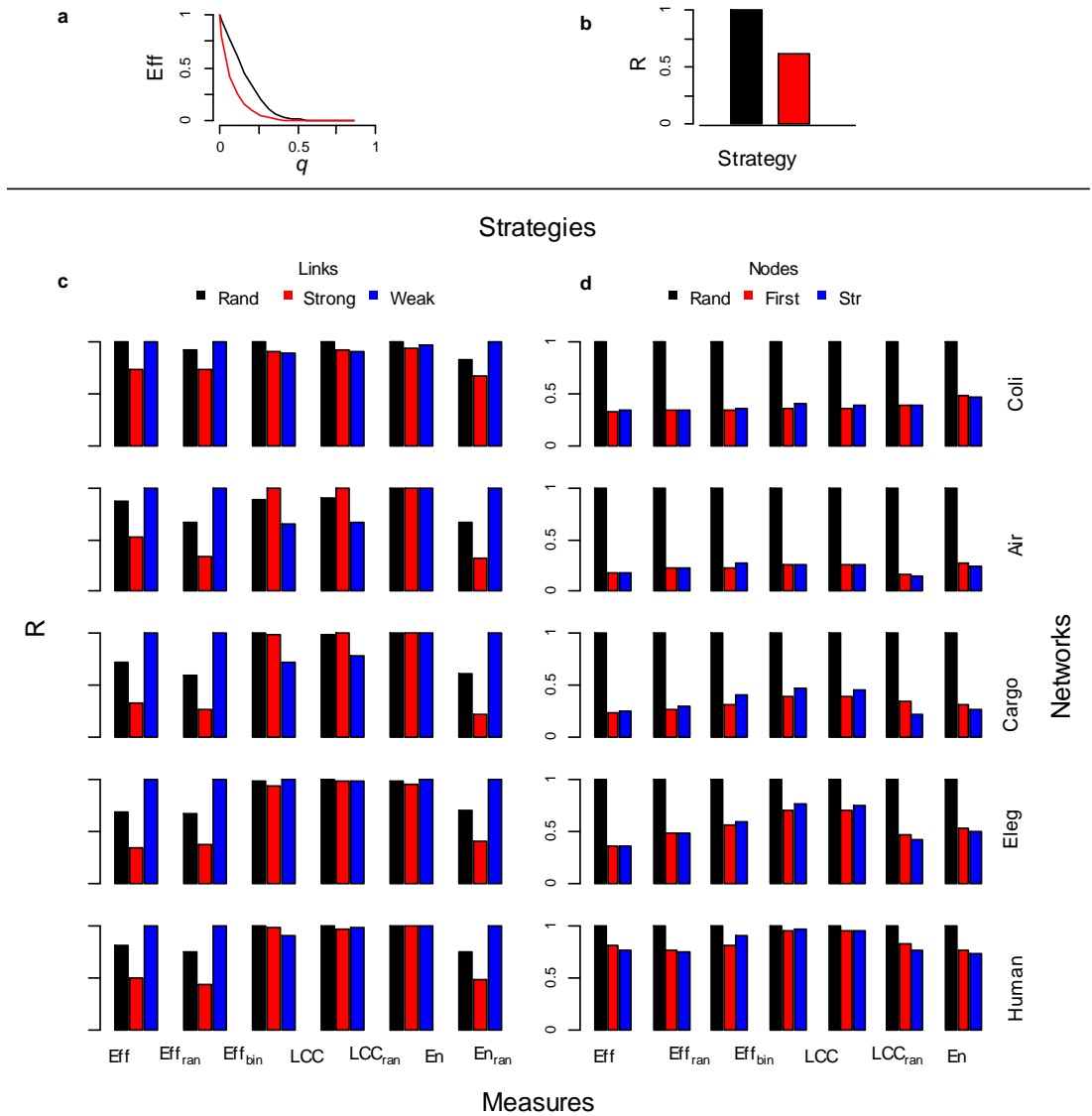

**Fig. 1**: **Robustness of the real-world complex weighted networks under different types of nodes-links removals**. The robustness of a high quality set of real-world complex weighted networks was analyzed under different nodes-links removals strategies and using different measures for the system functioning. **a**, Efficiency of the system (*Eff*) as function of the fraction of nodes or links removed ($q$) from the complex network. The network efficiency (*Eff*) decreases under nodes-links removal meaning that the global system functioning is negatively affected by the deletion of nodes or links; in this example the red strategy produced a sharper decrease in the network efficiency meaning that is more harmful than the black strategy to damage the system. **b**, example of the complex networks robustness (*R*) for the two strategies of nodes or links removal of the outcomes in (**a**). The robustness (*R*) of the removal strategy is the area below the curve produced by the removal strategy in (**a**). The robustness (*R*) produced by the removal strategy is normalized on the max value of the strategy robustness for that system functioning measurement; in this way we can easily compare the robustness of the network under different nodes-links removal strategies; here for example the black strategy produced higher network robustness (*R*) than the red strategy, thus the robustness of the strategies is normalized by the black function robustness value. **c,d**, The real-world networks robustness (*R*) under different nodes-links removal strategies for each system functioning measurement. **c**, links removal strategies. **d**, nodes removal strategies. *Eff* indicates the weighted system efficiency computed on the real-world networks; $Eff_{ran}$ is the weighted efficiency computed on the real-world network after the randomization of the links weight, i.e. the real link weights are randomly re-assigned on the network links. *LCC* indicates the largest connected cluster in in the network; $LCC_{ran}$ is the largest connected cluster measurements computed on the real-world network after the randomization of the links weight; *En* is the total energy-information in the network, i.e. the sum of the all links weight; $En_{ran}$ is the total energy computed on the real-world networks after the randomization of links weight.

**FIG. 2**

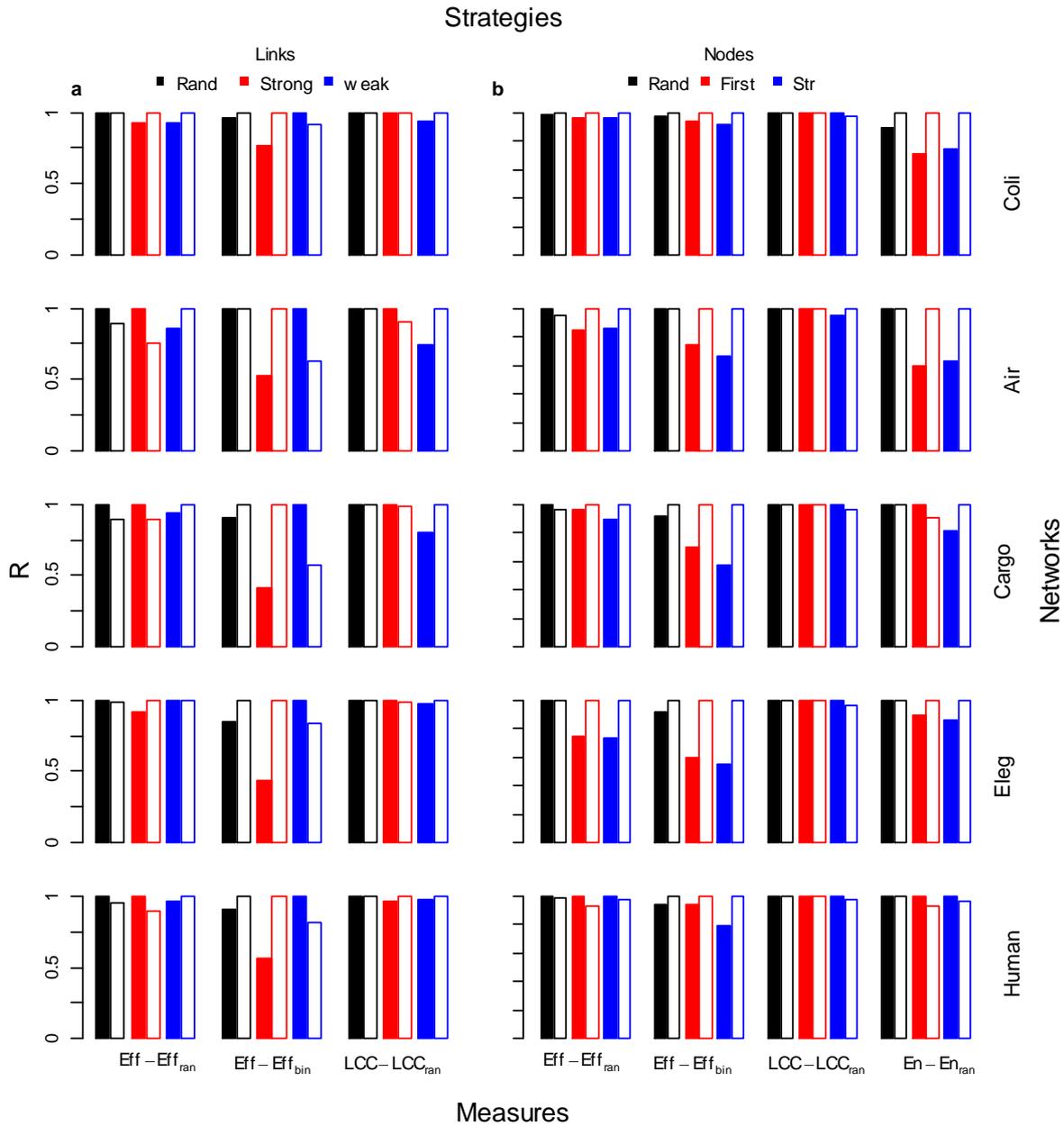

**Fig. 2**: **Robustness of the real-world complex weighted networks with real and randomized links weight.** Real-world complex weighted networks was analyzed comparing the robustness (*R*) of real and randomized links weight. In this figure we show the robustness (*R*) comparing outcomes of the real-world network with the randomized version of the same system. In the randomized network weights are reshuffle and randomly reassigned over the links; in this manner the binary-topological structure is maintained. The filled box indicates the robustness of the real-world system and the empty box of the same color contour shows the robustness of the randomized version of the same system. **a**, links removal strategies. **b**, nodes removal strategies. *Eff* indicates the weighted system efficiency computed on the real-world networks; *Eff$_{ran}$* is the weighted efficiency computed on the real-world network after the weights reshuffle, i.e. the real link weights are randomly re-assigned over the network. *LCC* indicates the largest connected cluster in in the network; *LCC$_{ran}$* is the largest connected cluster measures computed on the real-world network after the randomization of the weights; *En* is the total energy-information in the network, i.e. the sum of the all links weight; *En$_{ran}$* is the total energy computed on the real-world networks after the randomization of link weights. For the links removal strategy *En* is not plotted because the real and the randomization of link weights return the same hierarchy of links removal and thus the same outcomes of system energy decrease.

**FIG. 3**

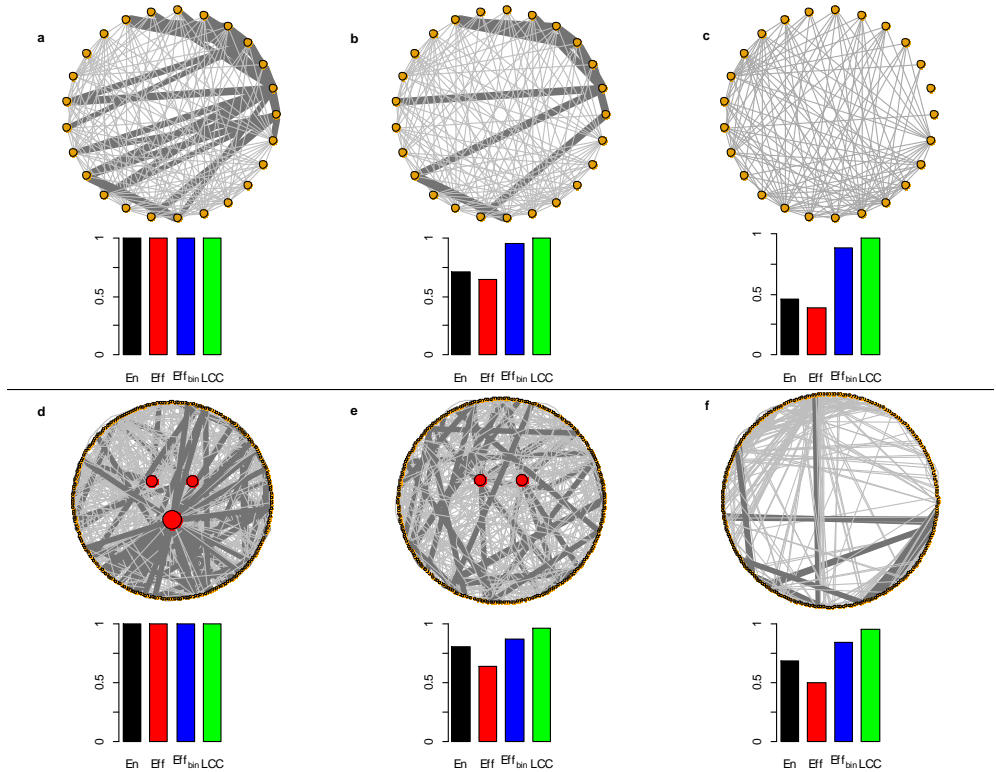

**Fig. 3**: **Abrupt functioning efficiency collapse and the 'connected but inefficient' network state.**

Step by step strong links removal simulation on a sub-network (N=25) of the US Airports real-world complex network. The total energy-information *En* and the weighted efficiency *Eff* readily decrease under the removal of links with higher weight. The binary network functioning parameters *Eff$_{bin}$* and *LCC* are constant under the removal of links with higher weight. The deletion of the strong links leaves the system in a 'connected but inefficient' network state. **a**, initial sub-network of 25 nodes and 153 links drew from US Airports complex weighted network. The weight of the links identifies the number of passengers flowing from the connected airports; the weights ranges from a minimum of 9 to maximum of 2253992 passengers per year. For the illustrative example we maintain only the strongest links with weight >$10^6$ (26 strongest links) and <380.000 passengers per year (127 weakest links). The dark grey links of major thickness are the strongest and thin soft grey links are the weakest. **b**, the sub-network after the removal of 16 links of highest weight (10% of the links). The system loses 29% of the total energy (*En*, black bar) and the 35% of the system functioning efficiency (*Eff*, red bar) but only the 4% of the binary efficiency (*Eff$_{bin}$*, blue bar) and no decrease in the largest connected cluster (*LCC*, green bar). **c**, the sub-network after the removal of 26 links of highest weight (17% of the links). Now the energy (*En*) is the 44% of the initial flowing energy meaning that the removal of only 17% of the links is able to roughly halves the number of passengers from the subset of airports. Further, the system functioning efficiency (*Eff*) collapses to 38% of the initial efficiency. At this final step where all the strongest links are removed, the network loses only the 12% of the binary efficiency (*Eff$_{bin}$*) and only one node is disconnected (4% *LCC* decrease). **d**, *C. Elegans* real-world complex weighted network representing nodes-neurons and the links connections number among them (link weights). The three red nodes of higher strength are outlined in the center where the others nodes are in the circle layout. **e**, *C. Elegans* network after the removal of the first node loses the 27% of the efficiency (*Eff*) and the 20% of the total-energy information *En*, but only 3% of the *LCC* and 2% of the *Eff$_{bin}$*. **f**, neuronal network of *C. Elegans* following the removal of the three main nodes is deprived of the links with higher weight (dark grey links) with a sharp decrease of the efficiency *Eff* (50%) and of the total-energy information *En* (32%). A multitude of weak interactions (soft grey links) holds the network still connected showing a minimal *LCC* (4%) and *Eff$_{bin}$* (5%) decrease.

**FIG. 4**

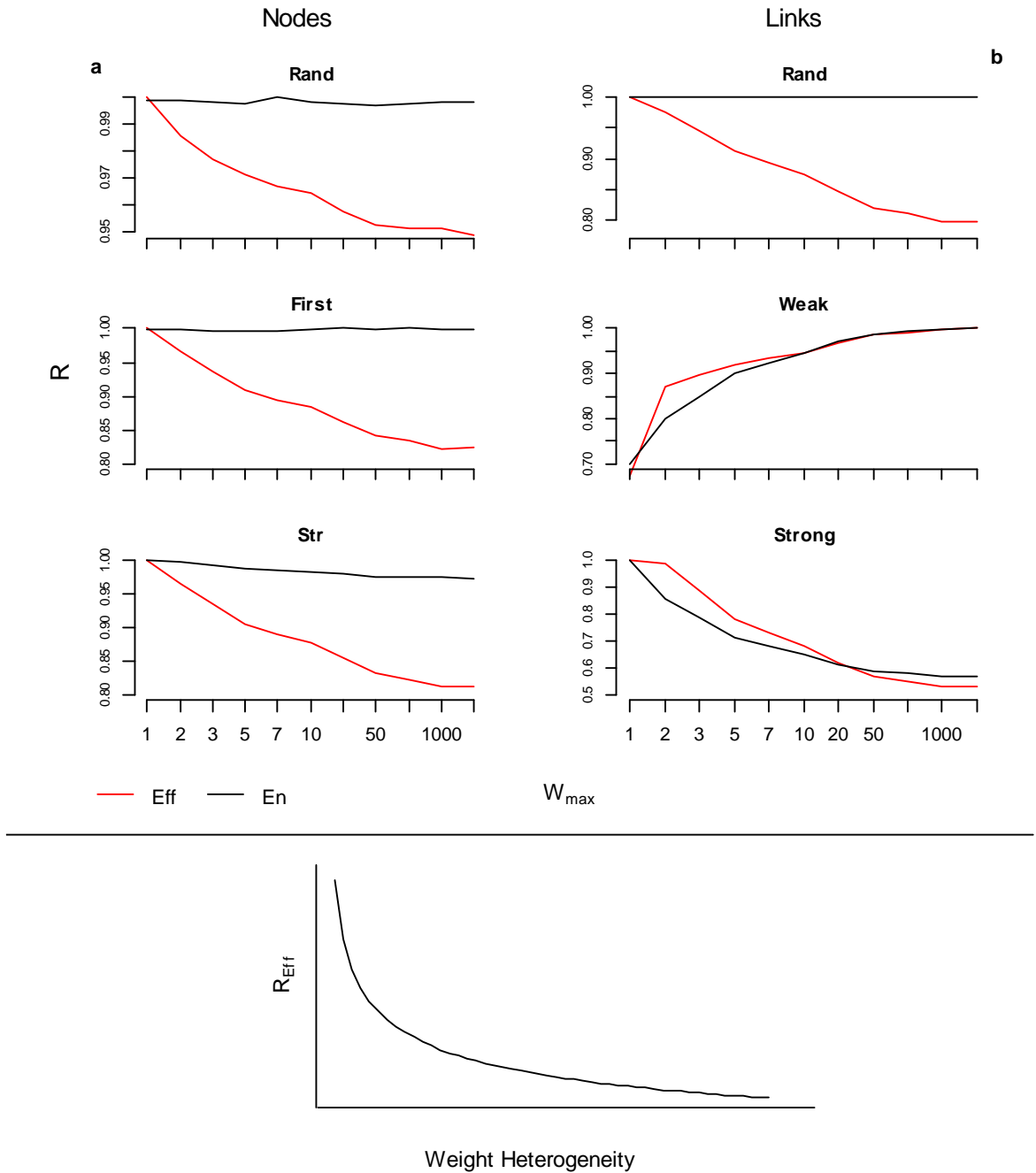

**Fig. 4**: **The real-world complex networks robustness of the efficiency functioning (*Eff*) decrease with links weight heterogeneity.**

The robustness of the efficiency *Eff* and total energy-information *En* of the system functioning under nodes and links removal strategies. We randomly assigned the weight of the links over the real-world topological structure of the networks under exam. Link weights are sorted from 2 values distribution (1,*Wmax*); the upper limit *Wmax* ranges in (1,10$^5$). For sake of example we depict the outcomes from the Cargo ship and the *E. coli* network. **a**, Cargo ship network under nodes removal strategies; **b**, *E. Coli* network under links removal strategies. In the bottom figure we exemplify the discovery outcomes: the robustness of the efficiency ($R_{Eff}$) is negatively correlated with the heterogeneity of links weight, i.e. increasing the variance in the weight of the links, the real-world complex network become more vulnerable to nodes-links removal, both selective and random.